\documentclass[prl,aps,twocolumn,amsfonts]{revtex4}

\usepackage{graphicx}
\usepackage{epsfig}

\def\qed{\leavevmode\unskip\penalty9999 \hbox{}\nobreak\hfill
     \quad\hbox{\leavevmode  \hbox to.77778em{%
               \hfil\vrule   \vbox to.675em%
               {\hrule width.6em\vfil\hrule}\vrule\hfil}}
     \par\vskip3pt}

\def\ibb #1{\leavevmode\hbox{\kern.3em\vrule
     height 1.5ex depth -.1ex width .4pt\kern-.3em\rm#1}}

\newcommand{\ba}[2]{\left(\begin{array}{#1}#2\end{array}\right)}
\newcommand{\tr}[1]{{\rm Tr}\left(#1\right)}
\newtheorem{theorem}{Theorem}
\newtheorem{lemma}{Lemma}

\begin{document}

\title{On the fidelity of mixed states of two qubits.}

\author{Frank Verstraete$^{ab}$ and Henri Verschelde$^a$\\
 $^a$Department of Mathematical Physics and Astronomy, Ghent University, Belgium\\
 $^b$Department of Electrical Engineering (SISTA), KULeuven,  Belgium}

\date{\today}

\begin{abstract}
We consider a single copy of a mixed state of two qubits and show
how its fidelity or maximal singlet fraction is related to the
entanglement measures concurrence and negativity. We characterize
the extreme points of the convex set of states with constant
fidelity, and use this to prove tight lower and upper bounds on
the fidelity for a given amount of entanglement.
\end{abstract}

 \maketitle

The concept of fidelity \cite{BDSW96}, also called maximal singlet
fraction \cite{HHH99}, is of central importance in the field of
quantum information theory. It is defined as as the maximal
overlap of the state with a maximally entangled state (ME)
\begin{equation}F(\rho)=\max_{|\psi\rangle={\rm
ME}}\langle\psi|\rho|\psi\rangle.\label{def1}\end{equation} An
explicit value for the fidelity has been derived by Horodecki
\cite{BHHH00}. If one considers the real $3\times 3$ matrix
$\tilde{R}=\tr{\rho\sigma_i\otimes\sigma_j}$ with
$\{\sigma_i,i=1..3\}$ the Pauli matrices, then
\[F(\rho)=\frac{1+\lambda_1+\lambda_2-{\rm Sgn}(\det(\tilde{R}))\lambda_3}{4}\]
with $\{\lambda_i\}$ the ordered singular values of $\tilde{R}$ and ${\rm Sgn}(\det(\tilde{R}))$ the sign of the determinant of $\tilde{R}$.\\

The concept of fidelity appears in the context of entanglement
distillation \cite{BBPSSW96,BDSW96} where it quantifies how close
a state is to a maximally entangled one, and in the context of
teleportation \cite{BBCJPW93} where it quantifies the quality of
the teleportation that can be achieved with the given state. Due
to the linearity and the convexity of the definition (\ref{def1}),
this measure has very nice properties that make it also possible
to derive upper bounds for the entanglement of distillation
\cite{Ra01}.\\

Despite the importance of the concept of fidelity, no rigorous
comparison appears to have been made before between the value of
the fidelity on one side and entanglement measures on the other
side. This paper aims at filling this gap and gives explicit tight
lower and upper bounds of the fidelity for given
concurrence \cite{Wo98} and negativity \cite{VW02}.\\

At first we will explicitly derive the possible range of values of
the fidelity in function of its concurrence \cite{Wo98} or
entanglement of formation. Next we show that the states that
minimize (maximize) the fidelity for given values of the
entanglement of formation are also extremal for given negativity
\cite{VW02}. Following \cite{EP99,VADD01}, we use the following
definition of negativity:
\[N(\rho)=\max\left(0,-2\lambda_{{\rm min}}(\rho^\Gamma)\right)\]
with $\lambda_{{\rm min}}$ the minimal eigenvalue of the partial
transpose of $\rho$ denoted as $\rho^\Gamma$.\\

\begin{theorem}
Given a mixed state of two qubits $\rho$ with negativity equal to
$N$ and concurrence equal to $C$, then its fidelity $F$ is bounded
above by
\[F\leq \frac{1+N}{2}\leq \frac{1+C}{2}.\] Moreover, the first
inequality becomes an equality iff $N=C$, and this condition is
equivalent to the condition that the eigenvector corresponding to
the negative eigenvalue of the partial transpose of $\rho$ is
maximally entangled.\label{theor1}
\end{theorem}
Proof: The fidelity of a state $\rho$ is given by {\small
\begin{eqnarray*}&&\max_{U_A,U_B\in SU(2)}\tr{(U_A\otimes
U_B)|\psi\rangle\langle\psi|(U_A\otimes U_B)^\dagger\rho}=\\
&&\frac{1}{2}\max_{U_A,U_B}\tr{\ba{cccc}{1&0&0&0\\0&0&1&0\\0&1&0&0\\0&0&0&1}(U_A\otimes
U_B^*)^\dagger\rho^{\Gamma}(U_A\otimes U_B^*)}
\end{eqnarray*}}
with $|\psi\rangle=(|00\rangle+|11\rangle)/\sqrt{2}$. An upper
bound is readily obtained by extending the maximization over all
unitaries instead of all local unitaries, and it follows that
$F\leq\tr{\rho^\Gamma}=(1+N)/2$. Equality is achieved iff the
eigenvector of $\rho^{T_\Gamma}$ corresponding to the negative
eigenvalue is maximally entangled. As shown in \cite{VADD01}, this
condition is exactly equivalent to the condition for $N$ to reach
its upper bound $C$, which ends the proof.\qed

Note that the upper bound is achieved for all pure states.\\

A more delicate and technical reasoning is needed to obtain a
tight lower bound on the fidelity. We will need the following
lemma:

\begin{lemma}\label{lemmaconvex}
Consider the density operator $\rho$ and the real $3\times 3$
matrix $\tilde{R}$ with coefficients
$\tilde{R}_{ij}=\tr{\rho\sigma_i\otimes \sigma_j}$ with $1\leq
i,j\leq 3$. Then $\rho$ is as a convex sum (i.e. mixture) of rank
$2$ density operators all having exactly the same coefficients
$\tilde{R}_{ij}$.
\end{lemma}
Proof: Consider the real $4\times 4$ matrix $R$ with coefficients
$R_{\alpha\beta}=\tr{\rho\sigma_\alpha\otimes\sigma_\beta}$,
parameterized as
\[R=\ba{cc}{1&\begin{array}{ccc}x_1&x_2&x_3\end{array}\\ \begin{array}{c}
y_1\\y_2\\y_3\end{array} & \tilde{R}}.\] If $\rho$ is full rank,
then a small perturbation on the values $\{x_i\},\{y_i\}$ will
still yield a full rank density operator. Consider a perturbation
on $x_1'=x_1+\epsilon$ and the corresponding $\rho'$. As the set
of density operators is compact, there will exist a lower bound
$lb<0$ and an upper bound $ub>0$ such that $\rho'$ is positive iff
$lb<\epsilon<ub$. Call $\rho_{lb}$, $\rho_{ub}$ the rank three
density operator obtained when $\epsilon=lb$ and $\epsilon=ub$
respectively. It is easy to see that
$\rho=(ub\rho_{lb}+lb\rho_{ub})/(lb+ub)$, such that it is proven
that a rank four density operator can always be written as a
convex sum of two rank three density operators with the same
corresponding $\tilde{R}$.\\
Consider now $\rho$ rank three and its associated "square root"
$\rho=XX^\dagger$ with $X$ a $4\times 3$ matrix. A small
perturbation of the form $\rho'=\rho+\epsilon XQX^\dagger$, with
$Q$ an arbitrary hermitian $3\times 3$ matrix $Q=\sum_{i=1}^9
q_iG_i$ and $G_i$ generators of U(3),  will still yield a state of
rank three. Moreover, there always exists a non-trivial $Q$ such
that $\tilde{R}$ is left unchanged by this perturbation. This is
indeed the case if  the following set of equations is fulfilled:
\[\sum_i q_i\tr{G_iX^\dagger(\sigma_\alpha\otimes
\sigma_\beta)X}=0\] for $(\alpha,\beta)=(0,0)$ and
$\alpha,\beta\geq 1$. It can easily be verified that this set of
10 equations only contains at most 8 independent ones irrespective
of the $4\times 3$ matrix $X$ , and as $Q$ has nine independent
parameters there always exists at least one non-trivial solution
to this set of homogeneous equations. A similar reasoning as in
the full rank case then implies that one can always tune
$\epsilon$ such that $\rho$ can be written as a convex sum of two
rank two density operators with the same $\tilde{R}$, which
concludes the proof.\qed

This lemma is interesting if one wants to maximize a convex
measure of a density operator (such as the entropy or an
entanglement monotone) under the constraint that the fidelity is
fixed: indeed, the fidelity is only a function of $\tilde{R}$, and
by the previous lemma we immediately know that states with maximal
entropy for given fidelity will have rank two. Note that exactly
the same reasoning applies when one wants to maximize a convex
measure under the constraint that the CHSH Bell-violation is fixed
\cite{VW02}, as this CHSH Bell-violation is also solely a function
of $\tilde{R}$. This is in exact correspondence with the results
derived in \cite{VW02}, where it was proven that the states
exhibiting the minimal amount of Bell violation for given
entanglement of formation are rank 2.

We are now ready to prove a tight lower bound on the fidelity:

\begin{theorem}
Given a mixed state of two qubits $\rho$ with concurrence equal to
$C$, then a tight lower bound for its fidelity $F$ is given by:
\[F\geq\max\left(\frac{1+C}{4},C\right).\]
\label{th2}
\end{theorem}
Proof: A direct consequence of lemma (\ref{lemmaconvex}) is that
to find states with minimal fidelity for given concurrence (i.e.
maximal concurrence for given fidelity), it is sufficient to look
at states of rank two. Consider therefore a rank 2 state $\rho$
and associated to it the real $4\times 4$ matrix $R$ with
coefficients $R_{\alpha\beta}=\tr{\sigma_\alpha\otimes
\sigma_\beta\rho}$. As shown in \cite{VDD01a,VDD02a,VADD01,VW02},
if $R$ is multiplied right and left by proper orthochronous
Lorentz transformations leaving the $(0,0)$-element equal to 1,
then a new state is obtained with the same concurrence. Moreover
the fidelity of a state $\rho$ is variationally defined as
\[F(\rho)=\min_{O_A,O_B\in
SO(3)}\tr{M\ba{cc}{1&0\\0&O_A}R\ba{cc}{1&0\\0&O_B^T}}\] with
$M={\rm diag}(1,-1,-1,-1)$ ($M$ is the representation of the
singlet in the R-picture). The minimal fidelity for given
concurrence can therefore be obtained by minimizing the following
constrained cost-function over all proper orthochronous Lorentz
transformations $L_1,L_2$:
\[K=\tr{ML_1RL_2^T}-\lambda\tr{L_1RL_2^T\ba{cccc}{1&0&0&0\\0&0&0&0\\0&0&0&0\\0&0&0&0}}.\]
Note that $\lambda$ is a Lagrange constraint. Without loss of
generality we can assume that the lower $3\times 3$ block
$\tilde{R}$ of $R$ is diagonal and of the form $\tilde{R}={\rm
diag}(-|s_1|,-|s_2|,-s_3)$ with $|s_1|\geq|s_2|\geq|s_3|$, as this
is precisely the form needed for maximizing the fidelity over all
local unitary operations. The cost-function $K$ can be
differentiated over $L_1,L_2$ by introducing the generators of the
Lorentz group (see e.g.\cite{VW02}), and this immediately yields
the optimality conditions $(\lambda=0, MRM=R^T)$ or $(\lambda=2,
R=R^T)$. Note however that the above argument breaks down in the
case that $|s_2|=-s_3$. Indeed, the fidelity cannot be
differentiated in this case as for example a perturbation of $s_3$
of the form $s'_3=s_3+\epsilon$ always leads to a perturbation of
the fidelity $F'=F+|\epsilon|$. In this case the conditions
$x_2=y_2,x_3=y_3$ or $x_2=-y_2,x_3=-y_3$ vanish, and if also
$|s_1|=|s_2|=-s_3$
there are no optimality conditions on $\{x_i,y_i\}$ left.\\
Let us first treat the case with $R$ symmetric and $s_1\geq
s_2\geq|s_3|$:
\[R=\ba{cccc}{1&x_1&x_2&x_3\\x_1&-s_1&0&0\\x_2&0&-s_2&0\\x_3&0&0&-s_3}.\]
The condition that $\rho$ corresponding to this state is rank 2
implies that all $3\times 3$ minors of $\rho$ are equal to zero.
Due to the conditions $s_1\geq s_2\geq|s_3|$, it can easily be
shown that a state of rank 2 (and not of rank 1!) is obtained iff
$x_1=0=x_2$ and $x_3=\pm\sqrt{(1-s_1)(1-s_2)}$ and
$1-s_1-s_2+s_3=0$. In this case the concurrence is equal to
$C=s_2$ and  the fidelity is given by $F=(s_1+s_2)/2$, and the
constraints become $1\geq s_1\geq s_2\geq(1-s_1)/2$ what implies
that $C\geq 1/3$. The minimal fidelity for given concurrence
occurs when $s_1=s_2$ and then $C=F$ which gives the lower bound
of the theorem in the
case of $C\geq 1/3$.\\
Let us now consider the case where $R=MR^TM$:
\[R=\ba{cccc}{1&x_1&x_2&x_3\\-x_1&-s_1&0&0\\-x_2&0&-s_2&0\\-x_3&0&0&-s_3}\]
with again $s_1\geq s_2\geq|s_3|$. Let us first note that, due to
the symmetry, $R$ has a Lorentz singular value decomposition
\cite{VDD01a} of the form $R=L_1\Sigma\tilde{M}ML_1^TM$ with
$\Sigma$ of the form ${\rm
diag}(|\sigma_0|,-|\sigma_1|,-|\sigma_2|,-|\sigma_3|)$ and
$\tilde{M}$ of the form ${\rm diag}(1,1,1,1)$ or ${\rm
diag}(1,-1,-1,1)$ or ${\rm diag}(1,-1,1,-1)$ or ${\rm
diag}(1,1,-1,-1)$. It follows that $\tr{R}=\tr{\Sigma\tilde{M}}$,
and due to the ordering of the Lorentz singular values,
$\tilde{M}$ has to be equal to the identity if $\tr{R}\leq 0$. But
$\tr{\Sigma}$ is just $-2C$ with $C$ the concurrence of the state,
and $\tr{R}=2-4F$ with $F$ the fidelity of the state. Therefore it
holds that $F=(1+C)/2$ if $\tr{R}\leq 0$ which corresponds to the
upper bound of the fidelity. Therefore only the case where
$\tr{R}>0$ has to be considered for finding lower bounds of the
fidelity. The condition that the state be rank 2 (and not rank 1)
immediately yields: $x_3=0$, $s_1+s_2-s_3=1$ and
$s_1+s_2=x_1^2/(1-s_2)+x_2^2/(1-s_1)$. If we only consider the
case with $\tr{R}>0$, it holds that $s_3<0$ and the inequality
constraints become $(1-s_1)/2\leq s_2\leq (1-s_1)\leq 2/3$. The
concurrence can again be calculated analytically and is given by
$C=(1-s_1-s_2-s_3)/2$, and it follows that $F=(1-C)/2$. Note that
the inequality constraints limit $C$ to be in the
interval $C\in\{0,1/3\}$, and so this bound is less stringent then the one stated in the theorem.\\
Let us now move to the degenerate case where $s_1> s_2=-s_3$:
\[R=\ba{cccc}{1& x_1 &x_2&
x_3\\y_1&-s_1&0&0\\y_2&0&-s_2&0\\y_3&0&0&s_2}.\] As $s_1>s_2$,
optimality requires $x_1=\pm y_1$. We first treat the case
$x_1=y_1$. Defining $\alpha=x_3/y_3$, a set of necessary and
sufficient conditions for being
rank 2 is given by: \begin{eqnarray*}0&=&x_1=y_1\\
0&=&x_2+\alpha y_2\\
0&=&\alpha^2-\alpha\frac{1-s_1}{s_2}+1\\
0&=&(x_2^2+x_3^2)-\alpha s_2(1+s_1).\end{eqnarray*} Under these
conditions the concurrence can again be calculated exactly and is
given by $C=s_2$, while the fidelity is given by $F=(1+s_1)/4$.
Note that the above set of equations only has a solution if
$(1-s_1)/2\geq s_2$, implying that $C\leq 1/3$. The fidelity will
now be minimal when $s_2=s_1$, and then $F=(1+C)/4$ which is the
second bound
stated in the theorem.\\
Let us now consider the degenerate case with $s_1> s_2=-s_3$ but
$x_1=-y_1$. The rank 2 condition implies that $s_1+2s_2=1$ and
$x_2=-y_2$ and $x_3=y_3$. Some straightforward algebra leads to
the condition
\[4\frac{1-s_1}{1+s_1}x_1^2+1-s_1^2-2x_2^2-2x_3^2=0.\]
Taking into account the constraints, the concurrence is again
given by $C=s_2=(1-s_1)/2$ and bounded above by $1/3$, while the
fidelity if equal to $F=(1+s_1)/4=(1-C)/2$. This bound always
exceeds the previously derived bound $F\geq (1+C)/4$ for $C\leq
1/3$, and is therefore useless.\\
It only remains to consider the case where $s_1=s_2=-s_3$:
\[R=\ba{cccc}{1&x_1&x_2&x_3\\y_1&-s_1&0&0\\y_2&0&-s_1&0\\y_3&0&0&s_1}.\]
Defining $\alpha=x_1/y_1$, the rank 2 constraint leads to the
following set of necessary and sufficient conditions: \begin{eqnarray*}0&=&x_2-\alpha y_2\\
0&=&x_3+\alpha y_3\\
0&=&s_1\alpha^2+\alpha(1-s_1)+s_1\\
0&=&\alpha(x_1^2+x_2^2+x_3^2)+s_1(1+s_1).\end{eqnarray*} The
inequality constraint reads $s_1\leq 1/3$, and the concurrence can
again be calculated exactly and is given by $C=s_1$. Therefore the
fidelity of these states obeys the relation $F=(1+C)/4$ for $C\leq
1/3$, which is the sharp lower bound.\qed

It might be interesting to note that all rank 2 states minimizing
the fidelity for given concurrence are quasi-distillable
\cite{HHH99,VDD01a} and have one separable and one entangled
eigenvector. More specifically, the states minimizing the fidelity
for $C\leq 1/3$ are, up to local unitaries, of the form
\[\rho=\ba{cccc}{\frac{1+C}{2}&0&0&0\\0&\frac{1-C+\sqrt{1-2C-3C^2}}{4}&-\frac{C}{2}&0\\0&-\frac{C}{2}&\frac{1-C-\sqrt{1-2C-3C^2}}{4}&0\\0&0&0&0},\]
and those for $C\geq 1/3$ of the form
\[\rho=\ba{cccc}{1-C&0&0&0\\0&C/2&-C/2&0\\0&-C/2&C/2&0\\0&0&0&0}.\]

\begin{figure}
\epsfig{file=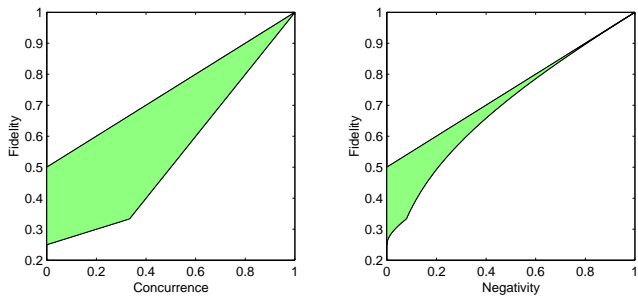,width=8.5cm} \caption{\label{fig:epsart}
Range of values of the fidelity for given concurrence and
negativity.} \label{fig1}
\end{figure}

Exactly the same states also minimize the fidelity for given
negativity. This leads to the following sharp bounds for the
fidelity versus negativity:
\begin{eqnarray*} F&\geq&\frac{1}{4}+\frac{1}{8}\left(N+\sqrt{5N^2+4N}\right)\\
F&\geq&\sqrt{2N(N+1)}-N\\
F&\leq& \frac{1+N}{2}.\end{eqnarray*} The first condition applies
when $N\leq (\sqrt{5}-2)/3$ and the second when $N\geq
(\sqrt{5}-2)/3$. A plot of these bounds is given in figure
(\ref{fig1}). One observes that the difference between the lower
bound and the upper bound in terms of the negativity becomes very
small ($\simeq \epsilon^2/16$) for large negativity
$N=1-\epsilon$. Moreover the fidelity is always larger then $1/2$
if the negativity exceeds $(\sqrt{2}-1)/2$.\\

In conclusion, we derived a tight upper bound for the fidelity for
given value of the concurrence and fidelity, and we identified all
states for which this upper bound is saturated. Next we have
characterized the extreme points of the convex set of states with
given fidelity, and this enabled us to derive tight lower bounds
on the fidelity for given amount of entanglement.

\bibliographystyle{unsrt}

\end{document}